\documentclass[12pt]{iopart}

\usepackage{graphicx}
\usepackage{subfigure}
\usepackage{iopams}
\usepackage{float}
\usepackage{amssymb}
\usepackage{amscd}
\usepackage{cite}
\usepackage{CJK}

\begin{document}

\title[W-J Su \emph{et al}]{Quantum information processing with nuclear spins mediated by a weak-mechanically
controlled electron spin}
\author{Wan-Jun Su}
\address{Department of Physics, Fuzhou University, Fuzhou
350002, China}
\address{Institute for Quantum Science and Technology, University of Calgary, Alberta T2N 1N4, Canada}
\ead{wanjunsu@fzu.edu.cn}
\author{Guang-Zheng Ye}
\address{Department of Physics, Fuzhou University, Fuzhou
350002, China}
\author{Ya-Dong Wu}
\address{Institute for Quantum Science and Technology, University of Calgary, Alberta T2N 1N4, Canada}
\address{QICI Quantum Information and Computation Initiative,Department of Computer Science, The University of Hong Kong, Pokfulam Road, Hong Kong}
\author{Zhen-Biao Yang}
\address{Department of Physics, Fuzhou University, Fuzhou
350002, China}
\author{Barry C. Sanders}
\address{Institute for Quantum Science and Technology, University of Calgary, Alberta T2N 1N4, Canada}

\begin{abstract}
We propose a scheme to achieve nuclear-nuclear indirect interactions mediated by a mechanically driven nitrogen-vacancy (NV) center in diamond. Here we demonstrate two-qubit entangling gates and quantum-state transfer between two carbon nuclei. When dipole-dipole interaction strength is much larger than the driving field strength, the scheme is robust against decoherence caused by coupling between the NV center (nuclear spins) and the environment. Conveniently, precise control of dipole coupling is not required so this scheme is insensitive to fluctuating positions of the nuclear spins and the NV center. Our scheme provides a general blueprint for multi-nuclear-spin gates and for multi-party communication.
\end{abstract}
\maketitle
\section{Introduction}

Solid-state quantum systems are advantageous for quantum-information applications due to
their inherent amenability to scaling~\cite{Nature101038}. Weakly coupled to the environment, nuclear spins have long coherence time in comparison with those of electron spins. Thus, nuclear spins are especially attrative for solid-state quantum storage~\cite{nphoton.2009.231} and for quantum gates~\cite{Nature07295}. Unfortunately, direct nuclear dipole-dipole interaction is negligible, which necessitates alternatives to couple nuclear spins.

NV center in diamond is a promising platform for highly sensitive nanoscale sensors~\cite{nature12072,NJP013020}. The NV center electronic
spin has exceptional quantum properties including high sensitivity to external
signals~\cite{Nature12016}, and a long spin coherence time~\cite{PRL015502}. Additionally, NV center spin states can be prepared and read out by optical pulses or microwave pulses at room temperature~\cite{nphys141}. These properties make NV center spin sensors an attractive candidate to detect nuclear-nuclear interactions~\cite{PRAPPL034031,PRL210502,PRAPPl044003}, and open up a way to exploit its applications to solid-state quantum information. Recently, Chen et al. used periodical resets of an NV center to protect a nuclear quantum sensor against decoherence and relaxation of the NV center~\cite{PhysRevLett.119.010801}. Via an extra auxiliary nitrogen-vacancy (NV) center electronic spin, we propose a scheme to control and mediate effective nuclear dipole-dipole indirect interaction.

Driving spin transitions of an NV center is the key to using NV center spins for nuclear-spin sensing. Except for optical and magnetic pluses, mechanical driving usually is applied to spin control. Significant progress in integrating NV centers with micro-electromechanical systems paves the way for spins coupled to mechanical resonators~\cite{nphys602,NJP045002,PRB064105,PRX041060,OpticsCom383.101}. Using mechanical driving, MacQuarrie et al. demonstrated direct spin-phonon interactions at room temperature as a means to drive magnetically forbidden spin transitions~\cite{PhysRevLett.111.227602}.

However, to achieve the nuclear-nuclear indirect interaction, two crucial challenges remain to be addressed:~(1) disorder in spin positioning and (2) small stress-coupling coefficient in driving spin transitions. Here we present an approach to overcome these challenges, to thereby achieve high-fidelity two-nuclear quantum gates and quantum-state transfer. In our scheme, an NV center is a mediator, coupling two nearby nuclear spins via dipole-dipole interactions. At the same time, mechanical~(stress) wave is used to drive the magnetically forbidden spin transition $\left|m_{s}=-1 \right\rangle\leftrightarrow \left|m_{s}=+1\right\rangle$ of the NV center spins. Our physical model requires that the driving-field Rabi frequency is sufficiently weak relative to the dipole-coupling strength. In experiments, normally, the stress-coupling coefficient is small so that large stress is required to produce a driving field~\cite{PhysRevLett.111.227602}. However, the small stress-coupling coefficient is helpful for our scheme, and large stress is not required, which increases the experimental feasibility. What's more, desired conditional manipulations do not require precise control of the dipole coupling. Hence, our scheme is robust against variations and uncertainties in the positions of the NV center and nuclear spins.

Paper is organized as follows. Section~\ref{sec:physical model} presents our physical model and the quantum dynamics of the model. In section~\ref{sec:quantum information processing} proposes a scheme to achieve entangling gates and quantum-state transfer based on this dynamics. We investigate fidelities versus parameter fluctuations~(scaled Rabi frequency, dipole-dipole coupling strength, etc.) via numerical simulations. Section~\ref{sec:Discussion} presents the discussions of our physical model at ambient conditions and further applications. Section~\ref{sec:Conclusions} is our conclusion.

\section{Physical model}

\label{sec:physical model}

\begin{figure}
\centering
\includegraphics[width=0.6\textwidth]{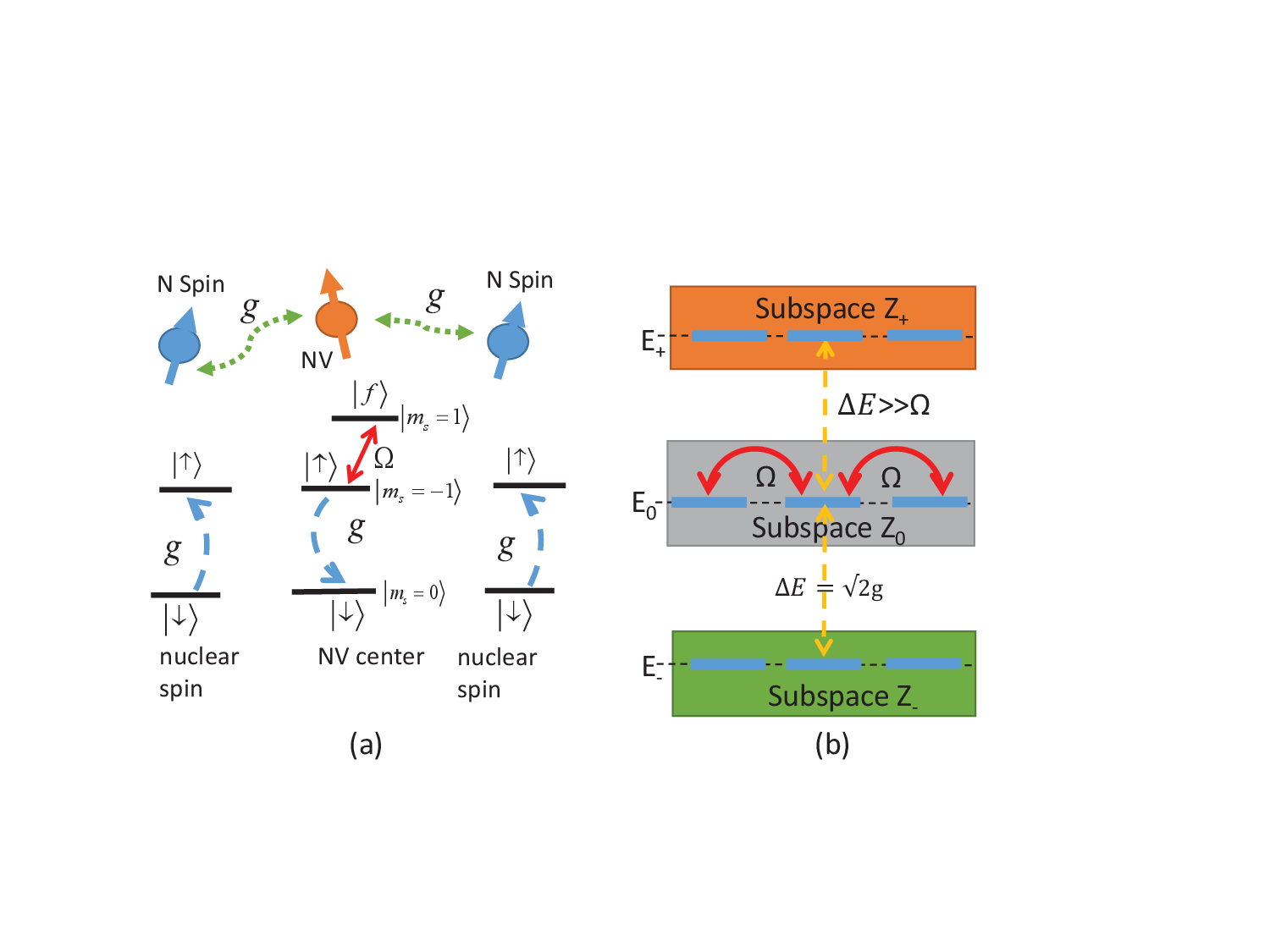}
\caption{(Color online)\label{Fig1}
(a)~Illustration of the basic principle of indirect interaction between two nuclear spins mediated by an NV center. The interaction between the NV center and nuclear spins occurs via coupling dipoles on the transition $  \left| \uparrow \right\rangle \leftrightarrow \left| \downarrow \right\rangle $, with coupling strength $g$. The transition $\left| f \right\rangle \leftrightarrow \left| \uparrow \right\rangle$  of the NV center is resonantly driven by mechanical (stress) wave of Rabi frequency $\Omega$. (b)~Schematic Hilbert subspaces. Due to the dipole-dipole interaction, the system energy is split into three energy levels \{~$Z_{+},Z_{0},Z_{-}$~\}. The energy split $\Delta E=\sqrt{2}g \gg \Omega$. The interaction between different states during the same subspace via the driving mechanical wave~(denoted by red dotted lines).}\label{Fig1}
\end{figure}

Our goal is to use the NV center as a mediator for indirect coherent interactions between two nuclear spins.  The ground state of the NV center is a spin-1 triplet state, with zero-field splitting $D=2.87$~GHz between the $\left| m_s=0 \right\rangle $ and $\left| m_s=\pm 1 \right\rangle $ states. By applying an external magnetic field along the NV center symmetry axis, one can lift the degeneracy of $\left| m_s=-1 \right\rangle$ and $\left| m_s=+1 \right\rangle$~\cite{nature09256}. For simplicity, we denote the following states: NV center spin states
\begin{eqnarray}\label{ }
\left| \downarrow \right\rangle_\mathrm{NV}&:=& \left| m_s=0 \right\rangle, \cr
\left| \uparrow \right\rangle_\mathrm{NV} &:=& \left| m_s=-1 \right\rangle, \cr
\left| f \right\rangle_\mathrm{NV} &:=& \left| m_s=1 \right\rangle,
\end{eqnarray}
and nuclear spin states
\begin{eqnarray}\label{ }
\left| \downarrow \right\rangle_\mathrm{N}&:=& \left| m_s=-\frac{1}{2},m_I=0 \right\rangle, \cr
\left| \uparrow \right\rangle_\mathrm{N} &:=& \left| m_s=\frac{1}{2},m_I=0 \right\rangle,
\end{eqnarray}

where $\mathrm{N}$ denotes the nuclear spin. The schematic setup is shown in Fig.~\ref{Fig1}(a). An NV center in diamond mediates two distinguishable nuclear spins. We assume that the interaction between the NV center and nuclear spins only occurs via dipole coupling on selected transitions $  \left| \uparrow \right\rangle \leftrightarrow \left| \downarrow \right\rangle $, whereas dipoles on other transitions do not interact with nuclear spins due to different frequencies or polarizations. When the energy between $\left| \uparrow \right\rangle_\mathrm{NV}$ and $\left| \downarrow \right\rangle_\mathrm{NV}$ of the NV center matches the transition frequency of the nuclear spin, the flip-flop process is most efficient. The dipole-coupling constant $g=(\mu_0 \gamma_e \gamma_\mathrm{N})/(4\pi r^{3})$ with $\mu_0$ the magnetic permeability, $\gamma_e$ and $\gamma_\mathrm{N}$ the gyromagnetic ratio of the electron spin and nuclear spin,respectively, and $r$ is the distance between the NV center and nuclear spin. In our scheme, we choose the dipole-coupling strength $g\sim2\pi\times2$~MHz, corresponding to the magnetic dipole-dipole interaction between $^\mathrm{14}$ NV electronic spin and $^\mathrm{13}$C nuclear spin~\cite{PhysRevB.79.075203}. A perpendicular stress couples $\left| f\right\rangle_\mathrm{NV}$ and $\left| \uparrow \right\rangle_\mathrm{NV} $ of the NV center, allowing a direct spin transition $ \left| f\right\rangle_\mathrm{NV}\leftrightarrow\left| \uparrow \right\rangle_\mathrm{NV} $ to be resonantly driven by a gigahertz-frequency mechanical (stress) wave. Here, the spin driving-field Rabi frequency $\Omega\sim2\pi\times 210$~kHz~\cite{PhysRevLett.111.227602}.

Under the rotating-wave approximation, the system can be described by the following
effective Hamiltonian in the interaction picture: ($\hbar=1$)
\begin{eqnarray}\label{Haminltonian 3}
H&=&H_\mathrm{DD}+H_\mathrm{drive}, \cr
H_\mathrm{DD}&=&\sum_{i=1,2}g_{i}\,\sigma_{\downarrow\uparrow}^\mathrm{NV}
\sigma_{i\uparrow\downarrow}^\mathrm{N}+\mathrm{H.c.},\cr
H_\mathrm{drive}&=&\Omega\left| f\right\rangle_\mathrm{NV}\langle\uparrow |+\mathrm{H.c.},
\end{eqnarray}
where $H_\mathrm{DD}$ describes the dipole-dipole interaction between the NV center spin and nuclear spins. $\sigma_{\alpha\beta}=\left| \alpha\right\rangle_\mathrm{NV}\langle\beta |$ are dipole operators and $\alpha,\beta\in \{\downarrow, \uparrow \}$. $H_\mathrm{drive}$ describes the stress waves driving the NV center,

Considering an initial system state $\left|\phi_0 \right\rangle$, the whole system evolutes in subspaces spanned \{ $\left|\phi_0 \right\rangle, \left|\phi_1 \right\rangle,\cdot\cdot\cdot \left|\phi_i \right\rangle$ \}, we reexpress $H_\mathrm{DD}$  as $\langle\phi_i|A\left|\phi_j \right\rangle$. The matrix A is real and can be diagonalized to $\Lambda$ via a transformation matrix $S$ such that $SAS^{\mathrm{T}}=\Lambda$. The degenerate eigenmodes come with energy $\lambda_j,j=1,2,3$, corresponding to eigenvectors $\left| \psi_k \right\rangle=\sum_{j} S_{k,j}\left| \phi_j \right\rangle$, where $k=1,\cdots n$. So that $H_\mathrm{DD}=\sum_n\lambda_nP_n$ with $P_k=\left| \psi_k\right\rangle\langle\psi_k |$. As shown in Fig.~\ref{Fig1}(b)., the eigenmodes is divided into three subspaces \{~$Z_+,Z_0,Z_-$~\} according to the corresponding energy $\lambda_{+},\lambda_{0},\lambda_{-}$, When the driving-field Rabi frequency is sufficiently weak relative to the dipole-coupling strength, the energy split $\Delta E=\sqrt{2}g \gg \Omega$, which prevents the transition between the different energy levels. The weak driving field only affects the quantum states during the same eigenmode-subspace, so that $H_\mathrm{drive}$ is considered as a perturbation term. If the initial system state $\left|\phi_0 \right\rangle$ is in the dark subspace $Z_{0}$, corresponding to the zero eigenenergy, the system evolves inside the dark subspace $Z_{0}$, which is the key in our physical model. The dynamical evolution process above is an analog of the strong continuous coupling in quantum Zeno dynamics ~\cite{JPA1751-8121-41-49-493001} or eigenmode-mediated unpolarized spin-chain state transfer~\cite{PhysRevA87.022306}. The effective system Hamiltonian is simplified to
\begin{eqnarray}\label{Equation 4}
H_\mathrm{eff}\simeq \sum_{n,\alpha,\beta}\lambda_nP_{n}+P_{n}^{\alpha}H_\mathrm{drive}P_{n}^{\beta},
\end{eqnarray}
where the projections $P_n=\left| \psi_n\right\rangle\langle\psi_n |$.

To model the system's dynamics, decoherence effects, such as the spontaneous decay of the NV center and nuclear spin with the corresponding relaxation rate $\Gamma_\mathrm{NV}$, $\Gamma_\mathrm{N}$, and the dephasing effect, are taken into account. We assume that the correlations of the spin bath degrees of freedom vanish on a short time span, which is negligible compared to the characteristic time scale of the system dynamics~\cite{QIP}. The memoryless~(Markovian) environment then enables one to derive simple equations of motion with the Lindblad formalism, leading to the following master equation~\cite{AO}
\begin{eqnarray}\label{Equation 5}
\frac{\mathrm{d}\rho}{\mathrm{d}t}=-\mathrm{i}[\rho,H]+\sum_{i=1,2}(\Gamma_{\mathrm{N}_i}+\Gamma_\mathrm{NV})L_\mathrm{S}[\rho]
+\sum_{i=1,2}(\gamma_{\mathrm{N}_i}+\gamma_\mathrm{NV})L_\mathrm{D}[\rho],
\end{eqnarray}
with the general form of $L[\rho]$
\begin{eqnarray}\label{6}
L_\mathrm{S}[\rho]=\sigma\rho\sigma^{+}-\frac{1}{2}(\sigma^{+}\sigma\rho-\rho\sigma^{+}\sigma),
\end{eqnarray}
and
\begin{eqnarray}\label{ }
L_\mathrm{D}[\rho]=\sigma_{z}\rho\sigma_{z}-\frac{1}{2}(\sigma_{z}\sigma_{z}
\rho-\rho\sigma^{+}_{z}\sigma_{z}),
\end{eqnarray}
corresponds to the relaxation and dephasing effect of electron spin or nuclear spins. Here, $\sigma$ is the dipole operator as $\sigma_{\alpha\beta}$ and
$\sigma_{z}=|\uparrow\rangle\langle\uparrow|-|\downarrow\rangle\langle\downarrow|$.
The spontaneous decay rate  $\Gamma_\mathrm{NV}\sim1$~KHz, $\Gamma_\mathrm{N}\sim 1$~KHz~\cite{Quant-ph:1904.08763}. The pure dephasing rate of NV spin and nuclear spin $\gamma_\mathrm{NV}\sim1$~KHz, $\gamma_\mathrm{N}\sim 1/3$~KHz~\cite{NaturePhysics408}. In the following numerical simulation, we considering $\Gamma=\Gamma_\mathrm{NV}=\Gamma_\mathrm{N}$ and $\gamma=\gamma_\mathrm{NV}=\gamma_\mathrm{N}$ for simply.

During the system dynamics, the system is initially in the state
$|\Psi_{0}\rangle$, and then evolves under Eq.~(\ref{Haminltonian 3}) for a choosing time, resulting in a final density $\rho$. The fidelity is defined as~\cite{UhlmannRMP273}
\begin{eqnarray}\label{ }
F:=\langle\Psi_{0}|\rho \left| \Psi_{0}\right\rangle,
\end{eqnarray}
where $\rho= \left| \Psi_{t}\right\rangle\langle\Psi_{t}|$. In the following, we show how to achieve the quantum entangling gates and quantum state transfer based on this model.

\section{Quantum information processing}
\label{sec:quantum information processing}

\subsection{two-qubit entangling gate}

Recently, Leonardo et al. proposed a scheme to realize
nonperturbative entangling gates between distant qubits using uniform cold atom chains~\cite{PhysRevLett.106.140501}. In
their work, an ideal mirror inverting dynamics generates a quantum gate $G$ between qubit
$A$ and qubit $B$, which reads
\begin{eqnarray}\label{Equation 9}
G|a\rangle|b\rangle=\mathrm{e}^{\mathrm{i}\phi_{ab}}|b\rangle|a\rangle,
\end{eqnarray}
where $a, b\in \{\downarrow, \uparrow \}$ in the computational basis. In the follows, we show how to achieve the two-qubit entangling gate based on our physical model.

We assume that quantum information is encoded in nuclear spin states $|\uparrow\rangle_\mathrm{N}$ and $|\downarrow\rangle_\mathrm{N}$, and the NV center
electron spin is an ancillary system, which is initially prepared in $|f\rangle_\mathrm{NV}$ state. For simplicity, we suppose there is only single excitation during the whole
system's evolution~\cite{PhysRevA80.012305}. Considering an initial system state $|\uparrow_1\uparrow_2\rangle_\mathrm{N}|f\rangle_\mathrm{NV}$, no dipole-dipole interaction takes effect. Time evolution of the system is within the subspace \{~$|\uparrow_1\uparrow_2\rangle_\mathrm{N}|f\rangle_\mathrm{NV}$, $|\uparrow_1\uparrow_2\rangle_\mathrm{N}|\uparrow\rangle_\mathrm{NV}$~\}. The driving field only causes a single-qubit rotation. After a single Rabi cycle, the nuclear spin state returns to its original state timing a $\pi$ phase shift.
\begin{eqnarray}\label{ }
 |\uparrow_1\uparrow_2\rangle_\mathrm{N}\rightarrow \mathrm{e}^{\mathrm{i}\Omega
 t}|\uparrow_1\uparrow_2\rangle_\mathrm{N}\rightarrow -|\uparrow_1\uparrow_2\rangle_\mathrm{N}.
\end{eqnarray}

When we consider an initial system state $|\uparrow_1\downarrow_2\rangle_\mathrm{N}|f\rangle_\mathrm{NV}$, the whole system
evolves in a single-excitation subspace $S_{1}$ spanned by

\begin{eqnarray}\label{ }
|\phi_1\rangle&=&|\uparrow_1\downarrow_2\rangle_\mathrm{N}|f\rangle_\mathrm{NV},\cr
|\phi_2\rangle&=&|\uparrow_1\downarrow_2\rangle_\mathrm{N}|\uparrow\rangle_\mathrm{NV},\cr
|\phi_3\rangle&=&|\uparrow_1\uparrow_2\rangle_\mathrm{N}|\downarrow\rangle_\mathrm{NV},\cr
|\phi_4\rangle&=&|\downarrow_1\uparrow_2\rangle_\mathrm{N}|\uparrow\rangle_\mathrm{NV},\cr
|\phi_5\rangle&=&|\downarrow_1\uparrow_2\rangle_\mathrm{N}|f\rangle_\mathrm{NV}.
\end{eqnarray}

The Hamiltonian of this subsystem reads
\begin{eqnarray}\label{ }
H_{\mathrm{total}}=\left(
\begin{array}{ccccc}
0 & 0 & 0 & 0 & 0 \\
0 & 0 & g_{2} & 0 & 0 \\
0 & g_{2} & 0 & g_{1} & 0 \\
0 & 0 & g_{1} & 0 & 0 \\
0 & 0 & 0 & 0 & 0 \\
\end{array}
\right)+\left(
\begin{array}{ccccc}
0 & \Omega & 0 & 0 & 0 \\
\Omega & 0 & 0 & 0 & 0 \\
0 & 0 & 0 & 0 & 0 \\
0 & 0 & 0 & 0 & \Omega \\
0 & 0 & 0 &\Omega & 0 \\
\end{array}
\right).
\end{eqnarray}
Then, setting $g_i=g$, under the condition $\Omega\ll g$, the whole Hilbert subspace is split into
three invariant subspaces according to the degeneracy of eigenvalues of $H_\mathrm{DD}$,

\begin{eqnarray}\label{ }
Z_{0}&=&\{ \left|\psi_{1}\right\rangle,\left|\psi_{2}\right\rangle, \left|\psi_{3}\right\rangle
\}, \cr
Z_{+}&=&\{\left|\psi_{4}\right\rangle\},\cr
Z_{-}&=&\{\left|\psi_{5}\right\rangle\},
\end{eqnarray}
with three corresponding eigenvalues $\lambda _1=0$, $\lambda _2=\sqrt{2}g$, $\lambda
_3=-\sqrt{2}g$, and where

\begin{eqnarray}\label{ }
\left|\psi_1\right\rangle&=& \left|\phi_1\right\rangle, \cr
\left|\psi_2\right\rangle&=&\frac{1}{\sqrt{2}}(-\left| \phi_2\right\rangle+\left|\phi_4\right\rangle), \cr
\left|\psi_3\right\rangle&=&\left|\phi_5\right\rangle, \cr
\left|\psi_4\right\rangle&=&\frac{1}{2}(\left| \phi_2\right\rangle+\sqrt{2}\left|\phi_3\right\rangle+\left| \phi_4\right\rangle), \cr
\left|\psi_5\right\rangle&=&\frac{1}{2}(\left| \phi_2\right\rangle-\sqrt{2}\left|\phi_3\right\rangle+\left| \phi_4\right\rangle).
\end{eqnarray}
Therefore, the Hamiltonian in Eq.~(\ref{Equation 4}) is approximately dominated by
\begin{eqnarray}\label{ }
H_\mathrm{eff} & \simeq & \sum_{n,\alpha,\beta}\lambda_n P_{n}+P_{n}^{\alpha}H_\mathrm{drive} P_{n}^{\beta} \cr
&=&\sqrt{2}g(|\psi_4\rangle\langle\psi_4|-\psi_5\rangle\langle\psi_5|)
+\frac{\Omega}{\sqrt{2}}(-\left|\psi_1 \right\rangle+\left|\psi_3 \right\rangle) \langle\psi_2|+\mathrm{H.c.}
\end{eqnarray}

Since the initial state is $|\Psi _{(0)}\rangle=|\phi_1\rangle$, the system will always evolve in the $Z_{0}$ subspace \{$ \left|\psi_{1}\right\rangle,\left|\psi_{2}\right\rangle, \left|\psi_{3}\right\rangle $ \} , and the Hamiltonian $H_\mathrm{eff}$ reduces to
\begin{eqnarray}\label{Equation 16}
H_\mathrm{e1}=\frac{\Omega}{\sqrt{2}}(-\left|\psi_1 \right\rangle+\left|\psi_3 \right\rangle) \langle\psi_2|+\mathrm{H.c.}
\end{eqnarray}
The general evolution of $H_\mathrm{e1}$ by solving the Schr$\ddot{o}$dinger equation with the interaction time $t$ is
\begin{eqnarray}\label{ }
|\Psi _{(t)}\rangle=\frac{1}{2}\bigg[(1+\cos\Omega
t)\left|\psi_{1}\right\rangle+(1-\cos\Omega t)\left|\psi_{3}\right\rangle+\sqrt{2}\mathrm{i}\sin\Omega t \left|\psi_{2}\right\rangle \bigg].
\end{eqnarray}
By choosing interaction time $t=\pi/\Omega$, the final state becomes
$|\Psi _{(t)}\rangle=|\downarrow_1\uparrow_2\rangle_\mathrm{N}|f\rangle_\mathrm{NV}$.Due to the symmetry of this system, an initial state of the system
$|\downarrow_1\uparrow_2\rangle_\mathrm{N}|f\rangle_\mathrm{NV}$, after a time $t=\pi/\Omega$, evolves to $|\uparrow_1\downarrow_2\rangle_\mathrm{N}|f\rangle_\mathrm{NV}$.

Similar to the step above, when the system is in initial state $|\downarrow_1\downarrow_2\rangle_\mathrm{N}|f\rangle_\mathrm{NV}$, the whole system
evolves in a single-excitation subspace $S_{2}$ spanned by:
\begin{eqnarray}\label{ }
|\phi_6\rangle&=&|\downarrow_1\downarrow_2\rangle_\mathrm{N}|f\rangle_\mathrm{NV},\cr
|\phi_7\rangle&=&|\downarrow_1\downarrow_2\rangle_\mathrm{N}|\uparrow\rangle_\mathrm{NV},\cr
|\phi_8\rangle&=&|\downarrow_1\uparrow_2\rangle_\mathrm{N}|\downarrow\rangle_\mathrm{NV},\cr
|\phi_9\rangle&=&|\uparrow_1\downarrow_2\rangle_\mathrm{N}|\downarrow\rangle_\mathrm{NV}.
\end{eqnarray}

Therefore, the invariant subspaces $S_{2}$ are
\begin{eqnarray}\label{ }
Z_{0}&=&\{\left|\phi _{6}\right\rangle,\left|\psi_{6}\right\rangle\}, \cr
Z _{+}&=&\{\left|\psi_{7}\right\rangle\}, \cr
Z _{-}&=&\{\left|\psi_{8}\right\rangle\},
\end{eqnarray}
the corresponding eigenvalues $\lambda _1=0$, $\lambda _2=\sqrt{2} g$, $\lambda
_3=-\sqrt{2} g$, where
\begin{eqnarray}\label{ }
\left|\psi_{6}\right\rangle &=&\left|\phi _{6}\right\rangle,\cr
\left|\psi_{7}\right\rangle &=&\frac{1}{\sqrt{2}}(-\left| \phi_{8}\right\rangle+\left|\phi_{9}\right\rangle), \cr
\left|\psi_{8}\right\rangle &=&\frac{1}{2}(\sqrt{2}\left| \phi_{7}\right\rangle+\left|\phi_{8}\right\rangle+\left| \phi_{9}\right\rangle), \cr
\left|\psi_{9}\right\rangle &=&\frac{1}{2}(-\sqrt{2}\left| \phi_{7}\right\rangle+\left|\phi_{8}\right\rangle+\left| \phi_{9}\right\rangle).
\end{eqnarray}
Because the initial state $\left|\psi _{6}\right\rangle$ is orthogonal to $\left|\psi_{7}\right\rangle$, $\left|\psi_{8}\right\rangle$ and
$\left|\psi_{9}\right\rangle$,~the effective Hamiltonian $H_\mathrm{e2}=0$. The state $|\downarrow_1\downarrow_2\rangle_\mathrm{N}|f\rangle_\mathrm{NV}$ remains unchanged
during system evolution.

Based on time evolution of our physics model, we can achieve the entangling gate of two distant
nuclear spins in diamond. For a duration of $T=\pi/\Omega$, the logical states of nuclear spins become

\begin{eqnarray}\label{ }
|\uparrow_1\uparrow_2\rangle_\mathrm{N}&\rightarrow&-|\uparrow_1\uparrow_2\rangle_\mathrm{N},\cr
|\uparrow_1\downarrow_2\rangle_\mathrm{N}&\rightarrow&|\downarrow_1\uparrow_2\rangle_\mathrm{N},\cr
|\downarrow_1\uparrow_2\rangle_\mathrm{N}&\rightarrow&|\uparrow_1\downarrow_2\rangle_\mathrm{N},\cr
|\downarrow_1\downarrow_2\rangle_\mathrm{N}&\rightarrow&|\downarrow_1\downarrow_2\rangle_\mathrm{N},
\end{eqnarray}

which corresponds to a two-qubit entangling gate between the distant nuclear spins. The
ancillary system (NV center), initially prepared in $|f\rangle_\mathrm{NV}$ state, is entangled
with the logical qubits during the gate operation, becoming once again disentangled by the
end of the operation.

\begin{figure}
\centering
\includegraphics[width=0.4\textwidth]{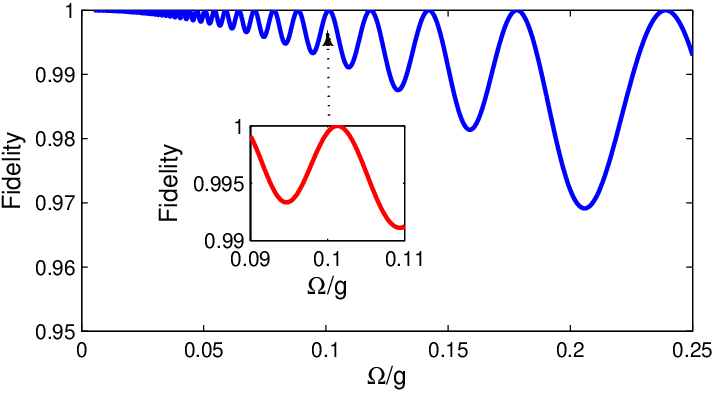}
\caption{(Color online)\label{Fig2}
The influence of the ratio $\Omega/g$ on the fidelity of the two-qubit entangling gate
under ideal conditions. $\Omega/g\in[0.005,0.25]$ and $\gamma_\mathrm{N}=\gamma_\mathrm{NV}=0$.}
\end{figure}

To verify the analytical results, we use numerical simulations to find the influences
of the interaction. Our model is valid when $\Omega\ll g$. Thus, we should consider the influence of the ratio $\Omega/g$ on the fidelity of the entangling gate. Fig.~\ref{Fig2} presents the fidelity as a function of $\Omega/g$ disregarding the decay. Not surprisingly, the fidelity decreases as the ratio $\Omega/g$ increases. The fidelity is $0.98$, even when $\Omega/g=0.15$. However, the interaction period $T=\pi/\Omega$ depends on $\Omega$, then a smaller ratio $\Omega /g$ results in longer operating time and increased decoherence. To balance the fidelity and the operating time, we choose $g\sim 2\pi\times 2 $~MHz and $\Omega\sim2\pi\times 210$~kHz to satisfy $\Omega/g=0.1$ in the following discussions.

\begin{figure}
\centering
\includegraphics[width=0.4\textwidth]{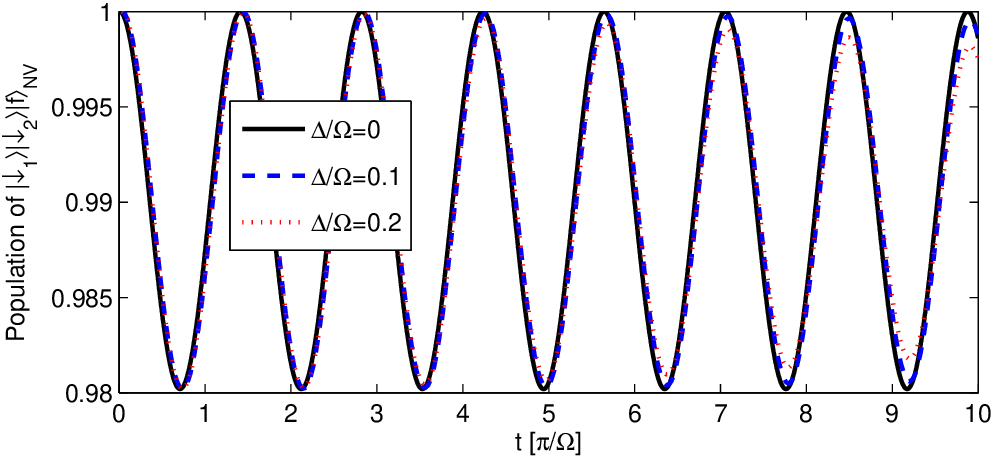}
\caption{(Color online)\label{Fig3}
The population of $|\downarrow_1\downarrow_2\rangle_\mathrm{N}|f\rangle_\mathrm{NV}$ as a function of the scaled time, considering the influence of detuning. The scaled ratio $\Delta/\Omega$ ranges from $0$ to $0.5$, $\gamma_\mathrm{N}=\gamma_\mathrm{NV}=0$, with $g\sim 2\pi\times 2 $~MHz and $\Omega\sim2\pi\times 210$~kHz.}
\end{figure}

The manipulation time depends on the stress-wave Rabi frequency. In our model, the
NV center resonantly interacts with the driving field of frequency $\omega_\mathrm{drive}$, and the corresponding transition energy of $|f\rangle_\mathrm{NV} \leftrightarrow |\uparrow\rangle_\mathrm{NV}$ is $\hbar\omega_{\uparrow f}$. However, under ambient conditions, the quantum dynamics is
affected by the off-resonant coupling, which will cause errors (phase shifts). We add fluctuating term in the driving field, so $H_\mathrm{drive}$ can be rewritten as
\begin{eqnarray}\label{ }
H_\mathrm{drive}=\Omega\cdot \mathrm{e}^{-\mathrm{i}\Delta t}\left|f\right\rangle_\mathrm{NV}\langle \uparrow |+\mathrm{H.c.},
\end{eqnarray}
where $\Delta :=\omega_{\uparrow f}-\omega_\mathrm{drive}$ describes the off-resonant coupling. The population of state $ |\downarrow_1\downarrow_2\rangle_\mathrm{N}|f\rangle_\mathrm{NV}$, as a function of the interaction time and scaled off-resonant coupling $\Delta/\Omega$, is shown in Fig.~\ref{Fig3}. When the requirement $\Delta/\Omega<0.2$ is met, small fluctuations of the population occur. Considering a small deviation of the resonant interaction between the driving field and NV center, we find that the average gate fidelity equals $0.995$ with $\Delta/\Omega=0.1$. The entangling gate is robust against fluctuations of the driving field, which is important to suppress errors.

\begin{figure}
\renewcommand\figurename{\small FIG.}
\centering
\subfigure[]{\includegraphics[width=0.4\textwidth]{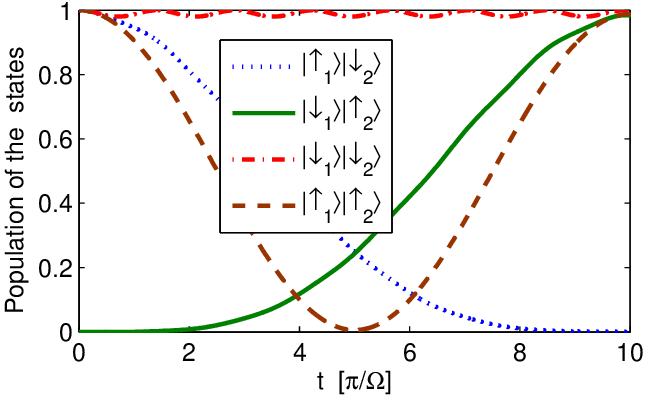}}
\subfigure[]{\includegraphics[width=0.4\textwidth]{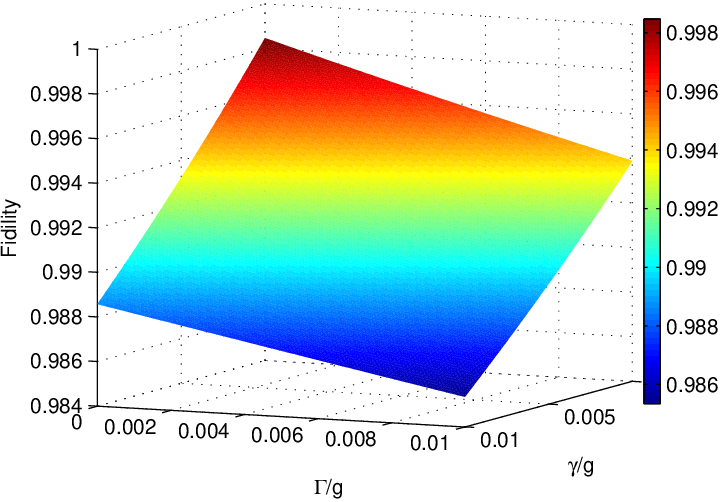}}
\caption{(Color online)\label{Fig4}
The influence of decoherence on realization of the two-qubit entangling gate, for $g\sim 2\pi\times 2 $~MHz and $\Omega\sim2\pi\times 210$~kHz. (a)
Dynamical evolution of the populations during the gate operation, with $\gamma_\mathrm{N}/g=\gamma_\mathrm{NV}/g=0.001$. (b) The fidelity of the entangling gate versus $\gamma_\mathrm{NV}/g$ and $\gamma_\mathrm{N}/g$.}
\end{figure}

The analysis above disregards the effects of decay. We now analyze how the gate operation
is affected by desired dissipation according to Eq.~(\ref{Equation 5}). Assuming scaled
spontaneous decay rates $\gamma_\mathrm{N}/g=\gamma_\mathrm{NV}/g=0.001$, Fig.~\ref{Fig4}(a) displays the dynamics of populations of nuclear spins states as a time function. At the end of interaction time $T$, $\rho_{\uparrow_{1}\uparrow_{2}}$ and $\rho_{\downarrow_{1}\uparrow_{2}}$ are about $0.985$. As shown in Fig.~\ref{Fig4}(b), the fidelity remains high ($>0.96$) with small scaled decay rates $\gamma_\mathrm{NV}/g$ and $\gamma_\mathrm{N}/g$, implying that the driving decouples the NV spin from the unwanted influence of environment.

\subsection{Quantum-state transfer}

Reliable quantum-state transfer (QST) between distant qubits has become a significant goal
of quantum physics research, owing to its potential application in a scalable quantum
information processing~\cite{Lloyd1569,PhysRevLett.90.057901,PhysRevLett.85.2392,NewJPhys.103041}. If the
system is initially in the state
\begin{eqnarray}\label{ }
|\Psi _0 \rangle=(\alpha |\downarrow_1\rangle_\mathrm{N}+\beta|\uparrow_1\rangle_\mathrm{N})
|\downarrow_2\rangle_\mathrm{N}|f\rangle_\mathrm{NV},
\end{eqnarray}
where $\alpha$, $\beta \in \mathbb{C}$, and $|\alpha|^{2}+|\beta|^{2}=1$. According to
quantum dynamics above, after an interaction time $t=\pi/\Omega$, the final system
state becomes
\begin{eqnarray}\label{ }
 |\Psi _\mathrm{F} \rangle=|\downarrow_1\rangle_\mathrm{N}(\alpha|\downarrow_2\rangle_\mathrm{N}+\beta|\uparrow_2\rangle_\mathrm{N})
|f\rangle_\mathrm{NV}.
\end{eqnarray}
The quantum information in nuclear spin 1 is transferred to nuclear spin 2.

To gain insight into the origin of this QST, we begin to consider a subset of Hilbert
subspaces of the system (see Fig.~\ref{Fig1}(b)). Dipole-dipole interaction causes a splitting of the system energy, with the corresponding energy eigenvalues $\lambda=0$,
$\lambda_{+}=\sqrt{2}g$, and $\lambda_{-}=-\sqrt{2}g$, respectively. Then the total Hilbert
space is split into three corresponding invariant subspaces
\begin{eqnarray}\label{ }
Z_{0}&=&\{\left|\Psi _{0}\right\rangle,\left|\Psi_\mathrm{F}\right\rangle,
|\Psi_\mathrm{I}\rangle\},\cr
Z _{+}&=&\{\left|\Psi_{+}\right\rangle_{1},\left|\Psi_{+}\right\rangle_{2} \}, \cr
Z _{-}&=&\{\left|\Psi_{-}\right\rangle_{1},\left|\Psi_{-}\right\rangle_{2} \},
\end{eqnarray}
where

\begin{eqnarray}\label{ }
|\Psi_\mathrm{I}\rangle&=& \frac{1}{2}(|\downarrow_1\uparrow_2\rangle_\mathrm{N}-|\uparrow_1\downarrow_2\rangle_\mathrm{N})(|\downarrow\rangle_\mathrm{NV}-|\uparrow\rangle_\mathrm{NV}),\cr
|\Psi_{\pm}\rangle_1&=& \frac{1}{2}(|\downarrow_1\uparrow_2\rangle_\mathrm{N}+|\uparrow_1\downarrow_2\rangle_\mathrm{N})|\downarrow\rangle_\mathrm{NV}\pm\sqrt{2}|\downarrow_1 \downarrow_2\rangle_\mathrm{N}|\uparrow\rangle_\mathrm{NV},\cr
|\Psi_{\pm}\rangle_2&=& \frac{1}{2}(|\downarrow_1\uparrow_2\rangle_\mathrm{N}+|\uparrow_1\downarrow_2\rangle_\mathrm{N})|\uparrow\rangle_\mathrm{NV}\pm\sqrt{2}|\uparrow_1 \uparrow_2\rangle_\mathrm{N}|\downarrow\rangle_\mathrm{NV}.
\end{eqnarray}
Large splitting of energy levels results in the difficulty of the transitions between different energy levels. However, the states belong to the same energy level interact with each other easily. If the initial state lies in the invariant subspace $Z_{0}$, the survival probability in $Z_{0}$ remains unity. A field drives the state transition from $\left|\Psi _{0}\right\rangle$ to $|\Psi_\mathrm{I}\rangle$, and then from $|\Psi_\mathrm{I}\rangle$ to $\left|\Psi_\mathrm{F}\right\rangle$. Mediated by the NV center, the quantum state is transferred between two distant nuclear spins,
\begin{eqnarray}\label{ }
&(\alpha |\downarrow_1\rangle_\mathrm{N}+\beta|\uparrow_1\rangle_\mathrm{N})
|\downarrow_2\rangle_\mathrm{N}|f\rangle_\mathrm{NV}     \cr
\rightarrow& (|\downarrow_1\uparrow_2\rangle_\mathrm{N}-|\uparrow_1\downarrow_2\rangle_\mathrm{N})(\alpha|\downarrow\rangle_\mathrm{NV}-\beta|\uparrow\rangle_\mathrm{NV}) \cr
\rightarrow& |\downarrow_1\rangle_\mathrm{N}(\alpha|\downarrow_2\rangle_\mathrm{N}+\beta|\uparrow_2\rangle_\mathrm{N})|f\rangle_\mathrm{NV}.
\end{eqnarray}
This process can be generalized to perform QST between any pair of multiple nuclear spins.

\begin{figure}
\centering
\includegraphics[width=0.4\textwidth]{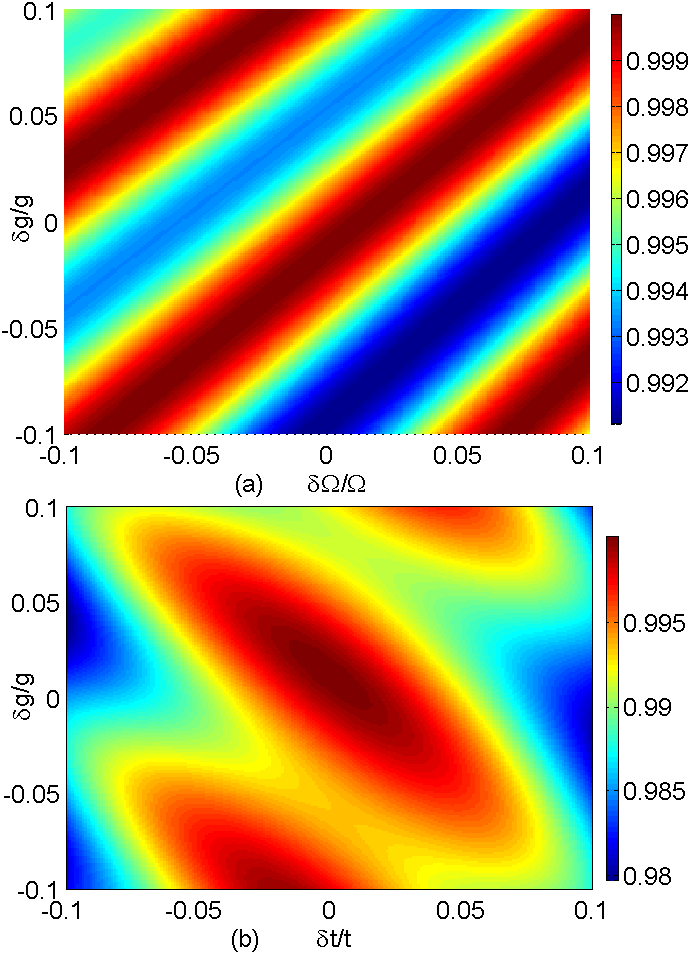}
\caption{(Color online)\label{Fig5}
The fluctuation of the interaction time, the coupling strength, and the Rabi frequency of
driving field influence the fidelity of QST, considering no spontaneous decay, for the
original $g \sim 2\pi\times 2 $~MHz and $\Omega \sim 2\pi \times 210$~kHz. (a)The
fidelity of QST versus $\delta g/g$ and $\delta\Omega / \Omega$. (b)The fidelity of QST versus $\delta g /g$ and $\delta t/t$ with the
original interaction time $T=\pi/\Omega$. }
\end{figure}

Now, we investigate the effect of systematic errors, which are caused by fixed fluctuations on the system parameters. For example, the fluctuation of driving-field Rabi frequency $\Omega$ can be assumed as a fixed value $\delta\Omega=\Omega^{'}-\Omega$ with $\Omega^{'}$ being the real value in experiment. $H_\mathrm{DD}$ in Eq.~(\ref{Haminltonian 3}) describes dipole-dipole interaction between the NV center spin and nuclear spins. A small uncertainty or variation in the separation $r$ leads to a corresponding change in $g$. A small fluctuation of the interaction time $t$ also affects QST. Thus, we consider three factors during the process of QST. Figure~\ref{Fig5} shows that small fluctuations of $g$, $\Omega$ and $t$ have little impact on the fidelity of QST, on condition that $\Omega \ll g$. Even a large fluctuation ($\delta g/g=0.1$, $\delta t/t=0.1$), the fidelity remains high $(\geq 0.98)$. The results of numerical simulation in fact demonstrate that QST between a nuclear-spin pair is robust against variations and uncertainty in the distance between the NV center and nuclear spins.

\begin{figure}
 \centering
 \includegraphics[width=0.4\textwidth]{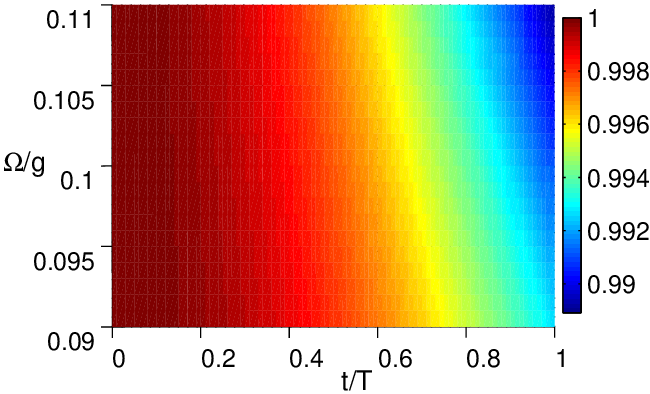}
\caption{(Color online)\label{Fig6} The survival probability in ground state $|0\rangle_\mathrm{NV}$ of NV center $P_0(t)$ versus the scaled time $t/T$ and $\Omega/g$, with $g \sim 2\pi \times 2$ MHz and the interaction time $T=\pi/\Omega$.}
\end{figure}

Considering the hybrid quantum devices based on NV centers in diamond, the dominant decoherence mechanism is photon emission via exciton decays of the NV center. If system is initially in the state $|f\rangle_\mathrm{NV}= |m_s=+1\rangle$, the survival probability reads
\begin{eqnarray}\label{ }
P_{0}(t)=\left[\frac{g^{2}+\Omega^{2}\cos (\sqrt{g^{2}+\Omega^{2}}t)}{g^{2}+\Omega^{2}}\right]^{2}.
\end{eqnarray}
As shown in Fig.~\ref{Fig6}, in regime $g\gg\Omega$, the survival probability in ground state $|f\rangle$ is about $1$. In our scheme, $\Delta E\gg\Omega$, weakly driving field keeps the NV center staying in the state $|m_s=+1\rangle_\mathrm{NV}$, which is belong to the zero engergy subspace. The probability in excited state is so small that the photon emission is suppressed.

\begin{figure}
\centering
\subfigure[]{\includegraphics[width=0.4\textwidth]{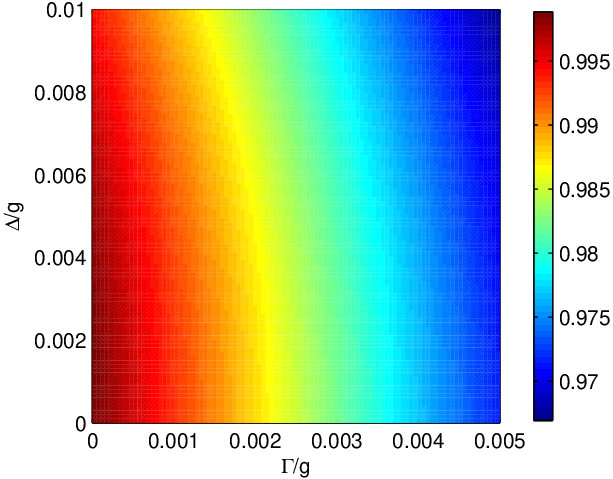}}
\subfigure[]{\includegraphics[width=0.4\textwidth]{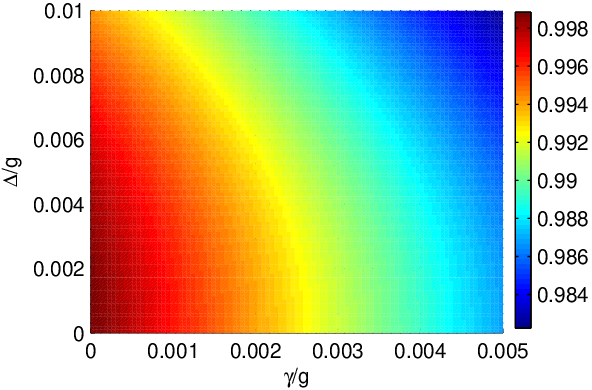}}
\caption{(Color online)\label{Fig7} The off-resonant coupling and the spontaneous decay of NV center and nuclear spin influence the fidelity of QST with $g\sim 2\pi\times 2$~MHz and $\Omega\sim2\pi\times 210$~kHz, for $t=\pi/\Omega$. (a) The fidelity of QST versus $\gamma_\mathrm{N}/g$ and $\Delta/g$. (b) The fidelity of QST versus $\gamma_\mathrm{NV}/g$ and $\Delta/g$. }
\end{figure}

Now, numerical simulations are used to show influences of spontaneous decay and off-resonant coupling. As shown in Fig.~\ref{Fig7}, when the spontaneous decay rate of NV center $\gamma_\mathrm{NV}/g\geq0.01$, with the scaled off-resonant strength $\Delta/g=0.01$, the fidelity $F\geq 0.94$. However, the fidelity $F\geq 0.97$, when the spontaneous decay rate of nuclear spin $\gamma_\mathrm{N}/g\geq0.01$. Thus, under the condition that $\Delta E\gg\Omega$, small off-resonant couplings and small decay rate of the NV center have little impact on the fidelity of QST.

\section{ Experimental Feasibility Discussion}
\label{sec:Discussion}

In this section, we explain the experimental realization of our model and present further applications. We first discuss how to prepare and manipulate the system spin states. Linearly polarized optical excitation preferentially pumps the NV center spin into ground state $|m_{s}=0\rangle_\mathrm{NV}$. The laser is then turned off and a magnetic adiabatic passage through the $|m_{s}=0\rangle_\mathrm{NV} \rightarrow |m_{s}=+1\rangle_\mathrm{NV}$ resonance robustly transfers the initialized spin population into the state $|m_{s}=+1\rangle_\mathrm{NV}$~\cite{PhysRevLett.111.227602}, which is the initial NV center spin state we need. Then, let magnetic-dipole coupling and mechanically driving take effect. A stress wave is turned on at a frequency $\omega_\mathrm{HBAR}$ corresponding to a resonance of the HBAR. The mechanical spin resonance $|m_{s}=+1\rangle_\mathrm{NV}\rightarrow |m_{s}=-1\rangle_\mathrm{NV}$ spin transition can be detected via optical pulses.

From the simulations above we can see that fidelity of the scheme is spoiled by dephasing. Therefore, implementing this proposal with high fidelity requires that the dephasing rate $\gamma_\mathrm{NV},\gamma_\mathrm{N} \ll g$. The relaxation time $T_{1}$ of an NV center can be achieved the order of seconds at low temperature around $4$~K~\cite{NatureCom11526}. The decoherence time $T_{2}$ can be prolonged to $15$ $\mu$s using a continuous dynamical decoupling of an NV center spin with a mechanical resonator~\cite{PhysRevB.92.224419}. In addition, in high purity diamond, $T_{2}$ of a single NV center is longer than $600$~$\mu$s at room temperature~\cite{Nature07279}. Nuclear spins have long relaxation and coherence time in comparison with those of electron spins in diamond. The relaxation time of single nuclear spin $T_{1}=(75\pm 20)$~$\mu$s at room temperature~\cite{PhysRevLett.97.087601}. The operating time required of entangling gates or QST is $t=\pi/\Omega\thickapprox 2.5$~$\mu$s, which is far less than $T_{1}$ and $T_{2}$ of the NV center (or nuclear spins). Therefore reasonable values of fidelity in our scheme can be anticipated.

Normally, driving NV electron spins independently of the nuclear spins is not easy. In our scheme, mechanical spin control of NV center and magnetic-dipolar coupling of nuclear-NV spins is a good alternative. However, there are also some challenges in experimental realization of our model. It is challenging to individually address one of the nuclear spins without affecting the other, no matter whether two nuclear spins have similar couplings to the NV center or not.

Our scheme can be extended to multi-qubit systems. An NV center spin is placed at the center of a polygon geometry with each vertex occupied by a nuclear spin. Based on our scheme, any two nuclear spins indirectly interact. Considered the polygon geometry a basic unit, many-body location can be achieved mediated by the interactions between distant NV centers~\cite{PhysRevLett.113.243002}. Furthermore, it may be used for simulating spin models with topological order~\cite{PhysRevLett.125.160503}.

\section{Summary}
\label{sec:Conclusions}

In summary, based on a weak driving field and dipole-dipole coupling, we present a protocol for the generation of entangling gates and quantum-state transfer between two separated nuclear spins mediated by an NV center in diamond. The system coherently evolves within the quantum dark subspaces. The results of numerical simulations show that
our protocol is robust against the fluctuations of external fields, and the uncertainty of the distance between the NV center and nuclear spins. This scheme also can be applied in biological probe. Nuclear spins in a single molecule were studied to learn structural information in chemical and biological processes~\cite{Science.283.1670}. Through dipole coupling to nuclear spins, an NV center can effectively act as a dipole ``antenna", detecting spins at different spatial locations. For example, NV centers can be used to detect the charge recombination rate in a radical pair reaction~\cite{PhysRevLett.118.200402}.

\section*{Acknowledgements}
The authors acknowledge J. P. Hadden and H.-Z. Wu for helpful discussions and suggestions.
This work is supported by the National Natural Science Foundation of China under Grant No. 11405031 and No.11875108, the National Natural
Science Foundation of Fujian Province china under Grant No.
2019J01219.

\end{document}